\documentclass[letterpaper, 10 pt, conference]{ieeeconf}

\IEEEoverridecommandlockouts
\usepackage{amsmath,amsfonts}
\usepackage{epsfig} 
\usepackage{times} 
\usepackage[ruled,linesnumbered]{algorithm2e}
\usepackage{array}
\usepackage[caption=false,font=normalsize,labelfont=sf,textfont=sf]{subfig}
\usepackage{textcomp}
\usepackage{stfloats}
\usepackage{url}
\usepackage{verbatim}
\usepackage{graphicx}
\usepackage{cite}
\usepackage{enumerate}
\usepackage{bm}
\usepackage{mathtools}
\usepackage{booktabs}
\usepackage{multirow}

\def\BibTeX{{\rm B\kern-.05em{\sc i\kern-.025em b}\kern-.08em
    T\kern-.1667em\lower.7ex\hbox{E}\kern-.125emX}}

\title{\LARGE \bf
Robust Integrated Priority and Speed Control based on Hierarchical Stochastic Optimization to Promote Bus Schedule Adherence along Signalized Arterial
}

\author{Shurui Guan, Keqiang Li, Haoyu Yang, Yihe Chen, Hanxiao Ren and Yugong Luo$^{\ast}$%
\thanks{This study was supported by the National Natural Science Foudation of China (NSFC) under grant number 52221005 and 5247120288.}
\thanks{Shurui Guan, Keqiang Li, Haoyu Yang, Yihe Chen, Hanxiao Ren and Yugong Luo are with the School of Vehicle and Mobility, Tsinghua University, Beijing 100084, China. Corresponding author: Yugong Luo (lyg@mail.tsinghua.edu.cn)}
\thanks{© 2025 IEEE. Personal use of this material is permitted.
  Permission from IEEE must be obtained for all other uses, in any current
  or future media, including reprinting/republishing this material for advertising
  or promotional purposes, creating new collective works, for resale or redistribution
  to servers or lists, or reuse of any copyrighted component of this work in other works.}
}

\begin{document}
\maketitle

\begin{abstract}

    In intelligent transportation systems (ITS), adaptive transit signal priority (TSP) and dynamic bus control systems have been independently developed to maintain efficient and reliable urban bus services. However, those two systems could potentially lead to conflicting decisions due to the lack of coordination. Although some studies explore the integrated control strategies along the arterial, they merely rely on signal replanning to address system uncertainties. Therefore, their performance severely deteriorates in real-world intersection settings, where abrupt signal timing variation is not always applicable in consideration of countdown timers and pedestrian signal design.
    In this study, we propose a robust integrated priority and speed control strategy based on hierarchical stochastic optimization to enhance bus schedule adherence along the arterial. In the proposed framework, the upper level ensures the coordination across intersections while the lower level handles uncertainties for each intersection with stochastic programming. Hence, the route-level system randomness is decomposed into a series of local problems that can be solved in parallel using sample average approximation (SAA). Simulation experiments are conducted under various scenarios with stochastic bus dwell time and different traffic demand. The results demonstrate that our approach significantly enhances bus punctuality and time headway equivalence without abrupt signal timing variation, with negative impacts on car delays limited to only 0.8\%-5.2\% as traffic demand increases.

\end{abstract}

\section{INTRODUCTION}

\subsection{Motivation}
Advanced public transit systems hold significant potential in alleviating traffic congestion. However, buses have been losing attractiveness to passengers stemming from inefficient and unreliable bus service, which is primarily induced by miscellaneous dynamic and stochastic operational factors, i.e. intersection delays and bus dwelling at stops. These unpredictable disturbances severely hinder buses from maintaining schedule adherence and even result in bunching. Fortunately, with the emergence of vehicle-to-infrastructure (V2I) and connected and automated vehicle (CAV) technologies, the traffic signal plans and buses behaviors could be adaptively controlled to mitigate random disturbances, engendering the potential for efficient and punctual bus operations.

Most existing research exclusively adopts either the perspective of intersection management or bus operators to improve bus service quality. For intersection management, adaptive transit signal priority (TSP) strategies are developed to prioritize buses by real-time traffic signal timing adjustment, which aims at minimizing bus delay with limited compromise to passenger vehicles. For bus operators, dynamic bus control strategy is proposed to regulate bus arrival time through bus holding or speed control, which emphasizes on disturbance resilience of the multi-bus system. However, ascribed to the lack of coordination, these two systems could potentially lead to conflicting objectives and non-optimal allocation of priority resources. Although a few studies have explored the integration of TSP and dynamic bus control using deterministic optimization, they entail frequent signal replanning to address stochastic perturbations, which will result in abrupt signal timing variation for current signal cycle. Such variation is not permitted in most intersections settings which are equipped with countdown timers or pedestrian signals, otherwise the numbers on countdown timers could change discontinuously and the “flash don't walk" phase of pedestrian signals will be difficult to design. Hence, it is of paramount advantage to develop an integrated priority and speed control strategy without reliance on abrupt signal timing variation, while still preserve high robustness under stochastic environment.

\subsection{Related Work}
From the perspective of dynamic bus control, most studies utilize bus holding strategies to regulate bus arrival time or time headway, represented by Daganzo~\cite{daganzoHeadwaybasedApproachEliminate2009}. In these studies~\cite{zhangTwowaylookingSelfequalizingHeadway2018, liangSelfadaptiveMethodEqualize2016}, the bus headway deviation is commonly designated as the error signal to design a feedback controller, so as to dynamically determine the bus dwelling time at each stop. Furthermore, optimization based methods are also adopted leveraging plentiful real-time information. In these studies~\cite{dengReduceBusBunching2020,bianOptimizationbasedSpeedcontrolMethod2020}, system performance within a finite time horizon is explicitly modeled and optimized within designed constraints, where real-time bus speed control could be additionally induced as an alternative or complement to holding strategies for better passenger experience. Recently, reinforcement learning based method has also garnered attention for its ability to effectively address environment randomness in a model-free manner~\cite{wangDynamicHoldingControl2020}. However, dynamic bus control strategies have to reactively respond to disturbances by extra dwelling or deceleration in constraint of traffic signals, resulting in adverse impact on bus travel efficiency.

From the perspective of adaptive TSP, extensive research has been conducted in isolated intersections. In these studies, strategies based on rules~\cite{leeRulebasedTransitSignal2013}, mathematical programming~\cite{huOptimizationModelBus2023}, and reinforcement learning~\cite{longDeepReinforcementLearning2022} have been proposed with in-depth consideration of various practical factors, such as queue effect~\cite{zengRealTimeTransitSignal2014}, bus-induced traffic bottlenecks~\cite{wuModelingOptimizingBus2020} and environmental uncertainties~\cite{liangOptimalControlImprove2022,liangRobustOptimalControl2024}. However, the priority time provided to buses is independently calculated rather than optimally allocated among multiple intersections, resulting in a wasteful priority resource allocation. Regarding arterial scenarios, existing studies mainly focus on maintaining signal coordination or green-wave progression as signal priority is granted~\cite{hanProgressionControlModel2022}. However, most of them solely pursue bus efficiency regardless of bus schedule adherence. Therefore, the priority might be dispensable for some buses and could exert adverse impact on route-level bus service reliability. Although Zeng~\cite{zengRouteBasedTransitSignal2021} proposes a route-based TSP strategy to address this gap by explicitly establishing the relationship of bus arrival time and multi-cycle signal timing plan, the integration with bus holding or speed control strategies is not considered.

Recently, leveraging V2I and CAV technologies, a few studies~\cite{chowMultiobjectiveOptimalControl2017,zimmermannBusTrajectoryOptimization2021,semanIntegratedHeadwayBus2020,bieDynamicHeadwayControl2020} have delved into the integrated design of TSP and bus speed control, which aims at leveraging the advantages of signal priority and bus control in a cooperative manner. Analogous to route-based TSP, most studies formulate the joint effect of TSP and speed guidance on bus arrival time across consecutive stops, then simultaneously optimize multi-intersection signal timing and cruise speed for all buses utilizing deterministic mathematical programming or optimal control. Nevertheless, due to the lack of explicit modeling for system randomness, frequent signal timing replanning is imperative to handle stochastic disturbances, which will inevitably result in abrupt signal timing variation. Hence, in real-world intersection settings where such variation is not always permitted, bus schedule adherence deteriorates significantly because of the model mismatch between deterministic models and stochastic environments.

\begin{figure*}[hb]
    \centering
    \resizebox{0.9\textwidth}{!}{\includegraphics{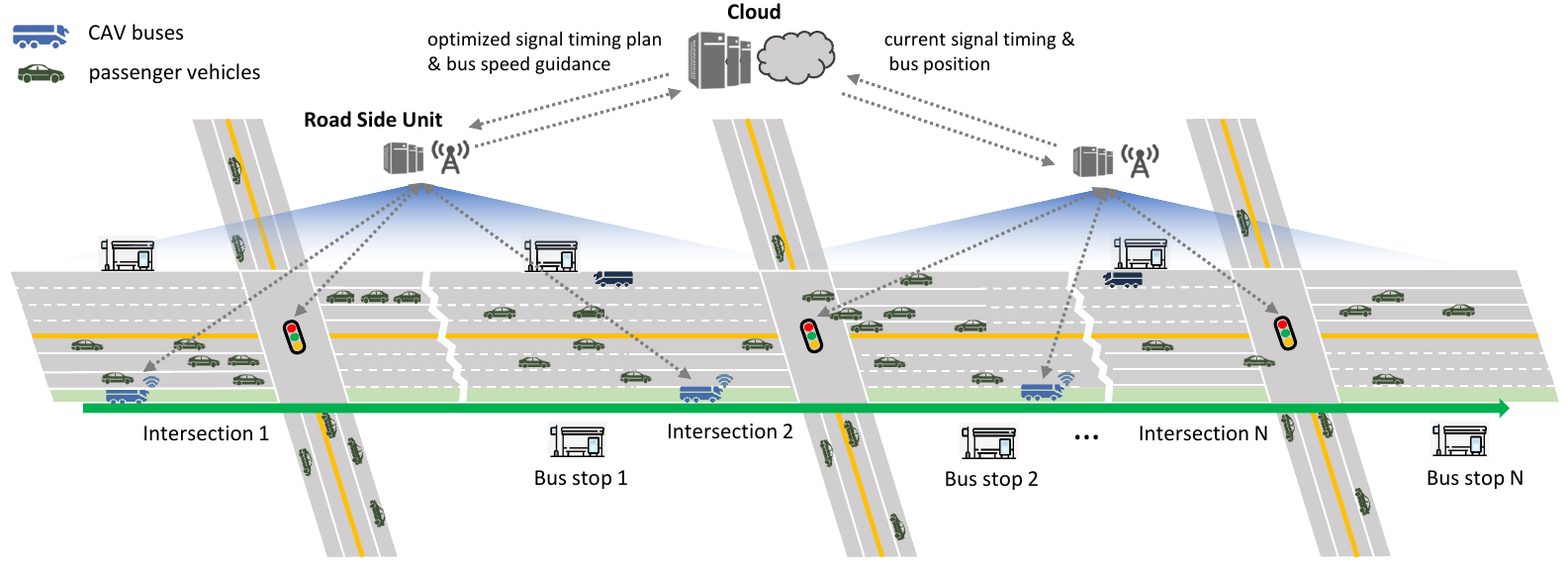}}
    \caption{The study scenario and control framework}
    \label{Fig:scenario}
\end{figure*}

\subsection{Contribution}
In this study, we focus on the integrated control strategies for promoting route-level bus schedule adherence under stochastic environments. A novel hierarchical stochastic programming-based integrated control strategy is proposed, which preserves high robustness to stochastic environmental factors without reliance on abrupt signal timing variation. The key contributions are outlined as follows:


\begin{itemize}
    \item [1)] For arterial-level coordination, we propose an integrated control strategy that jointly optimize multi-cycle traffic signal priority and multi-bus speed guidance in a global perspective. This formulation reduces the requirement of priority time by incorporating bus speed guidance, and optimally allocates multi-cycle signal timing across multiple intersections, thereby mitigating negative traffic impacts along the corridor.
    \item [2)] To address uncertainties, a stochastic programming-based strategy is developed for intersection-level signal priority and bus speed control. With explicit modeling of system randomness, a robust optimization scheme is obtained before the values of random variables are revealed, thus eliminating the reliance on abrupt signal timing variation. Hence, the bus schedule adherence can be significantly promoted under uncertainties in real-world intersection settings.
\end{itemize}

The subsequent sections are organized as follows. Section II introduces the study scenario and problem statement. Section III presents the hierarchical optimization framework and elaborates the optimization problem formulation for each level. Simulation are conducted and analyzed in Section IV, while Section V concludes the study.

\section{Problem Statement}

This study focuses on a bus route passing through the arterial, where each two adjacent bus stops are separated by a typical signalized intersection, as illustrated in Fig.~\ref{Fig:scenario}. All traffic signals and buses are equipped with V2X-enabled devices to support real-time data transmission with the central control system, such as an edge cloud. Buses are assumed to operate on bus lanes and could accurately execute received speed guidance commands. We assume a background signal timing plan, either optimized offline or updated periodically adapted to dynamic traffic flow, has already been provided, which could give satisfactory performance for overall traffic efficiency along the arterial.

Regarding environment randomness, there exist various uncertainties during bus operations, including dwelling, intersection queuing, surrounding vehicles behaviors et al. For clarity, we primarily consider stochastic dwell time as the representation of randomness, which is the most unpredictable and variable factor for dedicated-lane-aided bus operation, while uncertainties of other kinds could also be handled similarly. In addition, the distribution of bus dwell time in each stop is assumed known from historical statistics.

The objective of this study is to promote bus schedule adherence, i.e. the deviation between actual bus arrival time at stop with predefined schedule, under the premise of negligible impact on passenger vehicle delay. The proposed strategy jointly optimizes the timing plans for the next $K$ signal cycles at each intersection and the cruise speed for $N$ buses on each road segment downstream in real-time. Notably, once a signal cycle begins service, further modifications to signal timing of this cycle are prohibited to preserve continuous countdown display of traffic lights, with the propose of avoiding driver confusion and potential risk for cross-street pedestrians.

\section{Methodology}

\subsection{Hierarchical Optimization Framework}
\begin{figure}[h]
    \vspace{-2mm}
    \centering
    \includegraphics[width=\linewidth]{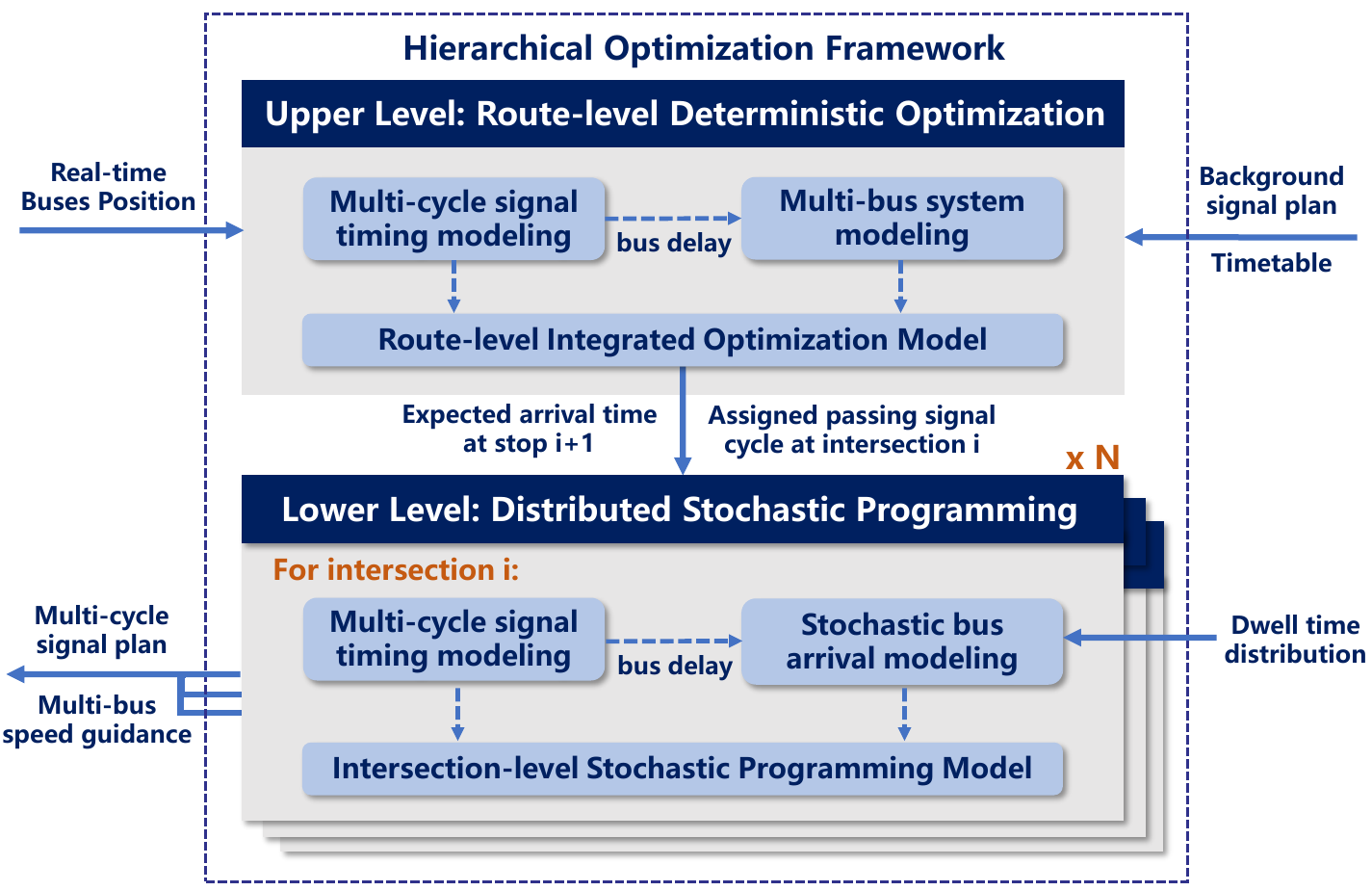}
    \caption{Hierarchical optimization framework}
    \label{Fig:framework}
\end{figure}

Given the large-scale nature of the joint optimization problem incorporating multi-intersection signal timing and multi-bus movements control in a long time horizon, explicitly modeling stochastic dwelling process in arterial-level would lead to intractable complexity. To address this issue, we propose a hierarchical optimization framework that decouples the arterial-level coordination of multiple traffic signals and buses from the management of stochastic disturbances, as illustrated in Fig.~\ref{Fig:framework}. Although theoretical optimality is sacrificed to some extent, the framework strikes a good balance between computational efficiency and solution quality.

The proposed framework consists of two hierarchical optimization models. The upper model optimizes the overall bus schedule adherence on route-level in a deterministic formulation. It takes real-time bus locations and intersection traffic conditions as input, outputting the expected arrival time at each stop and the assigned passing signal cycle at each intersection for all buses, based on pre-defined background signal timing plan and bus timetable. Although the randomness is neglected in upper model formulation, the coordination between signalized intersections and multiple buses are holistically modeled, ensuring the route-level optimization in a global perspective. 

The lower model is formulated based on stochastic programming and can be executed distributedly for each intersection, aiming at generating robust signal timing plan and bus speed reference under the guidance of upper-level outputs. Given the assigned bus-passing signal cycle and next-stop arrival time scheduled by upper model, a stochastic programming problem is formulated for each intersection, where the expectation of bus arrival time deviation is minimized under explicit modeling of stochastic bus dwell time. The stochastic formulation of lower model could eliminate the model mismatch problem, thereby enhance the effectiveness and robustness of bus schedule adherence. 

Furthermore, a time-triggered rolling horizon mechanism is employed in response to the ever-changing environment. When the algorithm is triggered, the upper model is solved at first, with its outputs dispatched to each lower model individually. Thereafter, all lower models are solved in parallel for each intersection. Finally, the control commands are generated by combining all lower model outputs and executed by all traffic signals and buses.

\subsection{Upper Level: Route-level Deterministic Optimization}
For upper model, our approach extends the idea of R-TSP strategy proposed by~\cite{zengRouteBasedTransitSignal2021} with the incorporation of bus speed guidance. The objective function consists of two weighted components. The first term represents the minimization of total bus arrival time deviations, while the second term aims at minimizing total green time compression compared to background signal plan, as defined in (\ref{Eq:upp_obj}).
\begin{equation}
    \label{Eq:upp_obj}
    \begin{aligned}
        \min J=\sum_{i=1}^{I} (\sum_{n=1}^{N} w_b\left| t_{ni}^{\text{arr}}-t_{ni}^{\text{opt}} \right| + \sum_{j \in P} \sum_{k=1}^{K} w_c(G_{ij}^{\text{opt}}-g_{ijk}))
    \end{aligned}
\end{equation}
where $t^{arr}$ and $t^{opt}$ denote the actual and timetable-scheduled bus arrival time, $g$ and $G^{opt}$ denote the green split (duration) to be planned and that from background signal plan. The subscript $n, i, j, k$ represent the index of bus, intersection, signal phase and signal cycle respectively.

To explicitly express the objective function~\ref{Eq:upp_obj} with decision variables, two key components have to be considered: (a) modeling of multi-cycle traffic signal timing and (b) modeling of bus movement along arterial.

For signal timing and phase modeling, we adopt the standard dual-ring structure to define phase relationships within and between signal cycles, mathematically described by~\ref{Eq:upp_tls_inK}-\ref{Eq:upp_tls_barr}. The start time and duration for phase $j$ at intersection $i$ in cycle $k$ are denoted as $t_{ijk}$ and $g_{ijk}$ respectively.
\begin{equation}
    \label{Eq:upp_tls_inK}
    \begin{aligned}
        t_{ij'k}=t_{ijk}+g_{ijk}+Y, \quad j \in P_{i} \backslash P_{i}^{\text{last}},
        \forall i,k
    \end{aligned}
\end{equation}
\begin{equation}
    \label{Eq:upp_tls_interK}
    \begin{aligned}
        t_{ij,k+1}=t_{ijk}+g_{ijk}+Y, \quad k=1, j \in P_{i}^{\text{first}}, \forall i
    \end{aligned}
\end{equation}
\begin{equation}
    \label{Eq:upp_tls_barr}
    \begin{aligned}
        t_{ijk}=t_{ij'k}, \quad j \neq j' \in P_{i}^{\text{barrier}}, \forall i,k
    \end{aligned}
\end{equation}
where $Y$ is the yellow time, $P_{i}$ represents the set of phases with superscript $\text{first}$, $\text{last}$ and $\text{barrier}$ referring to different kinds of phases. Since the start time of current phase $t_{ij\tilde{k}}$ is known, we can determine the boundary condition of (\ref{Eq:upp_tls_inK})-(\ref{Eq:upp_tls_barr}).

Moreover, a minimal green time constraint is introduced based on phase saturation to ensure queue clearance at each green phase on the average sense, formulated as (\ref{Eq:upp_tls_minG}).
\begin{equation}
    \label{Eq:upp_tls_minG}
    \begin{aligned}
        g_{ijk} \geq \max \{ \frac{V_{ij}C}{S_{ij}X_c}, G^{\text{min}} \}, \quad \forall i,j,k
    \end{aligned}
\end{equation}
where $V_{ij}$ and $S_{ij}$ denote the traffic volume and saturation rate, $C$ is the common signal cycle length, $X_c$ is the critical degree of saturation that a phase should not exceed, and $G^{\text{min}}$ represents minimal green time.

Furthermore, constraints~(\ref{Eq:upp_tls_lenKC})-(\ref{Eq:upp_tls_yp}) are introduced to maintain signal coordination across multiple intersections:
\begin{equation}
    \label{Eq:upp_tls_lenKC}
    \begin{aligned}
        t_{ijk}+g_{ijk}+Y=KC, \quad k=K, j \in P_{i}^{\text{last}}, \forall i
    \end{aligned}
\end{equation}
\begin{equation}
    \label{Eq:upp_tls_yp}
    \begin{aligned}
        \left| t_{ijk}-t_{i'jk} \right| \leq \left| t_{ij}^{\text{opt}}-t_{i'j}^{\text{opt}} \right| + \delta_{c}, \quad j \in P_{i}^{\text{coord}}, \forall i,k
    \end{aligned}
\end{equation}
(\ref{Eq:upp_tls_lenKC}) ensures that total cycle length over the planning horizon remains unchanged, thereby prevents continual shifting of the offset. (\ref{Eq:upp_tls_yp}) limits the variation of the green-wave band for passenger vehicles to a certain threshold $\delta_c$, which could preserves signal coordination thus mitigate negative traffic impacts on the main street.

For bus movement modeling, without loss of generality, we divide the whole bus route into a series of segments, which consists of upstream stop $i$, intersection $i$ and downstream stop $i+1$. At each segment, the bus movement can be further decomposed into four processes, i.e. dwelling at stop $i$, approaching intersection $i$, delay at intersection $i$, and approaching stop $i+1$, described by (\ref{Eq:upp_bus_app})-(\ref{Eq:upp_bus_dep}):
\begin{equation}
    \label{Eq:upp_bus_app}
    \begin{aligned}
        r_{ni}=t_{ni}^{\text{arr}}+T_{ni}^{\text{app}}+T_{ni}^{\text{st}}, \quad \forall n,i
    \end{aligned}
\end{equation}
\begin{equation}
    \label{Eq:upp_bus_dep}
    \begin{aligned}
        t_{n,i+1}^{\text{arr}}=r_{ni}+d_{ni}+T_{ni}^{\text{dep}}, \quad \forall n,i
    \end{aligned}
\end{equation}
where $r_{ni}$ is the arrival time at intersection $i$ for bus $n$, $T^{\text{st}}$, $T^{\text{app}}$, $d_{ni}$ and $T^{\text{dep}}$ denote the time duration of the four processes respectively. Utilizing bus speed adjustment, $T^{\text{app}}$ and $T^{\text{dep}}$ are modifiable, thus serve as decision variables constrained by speed limits $v^{\text{max}}$, as shown in (\ref{Eq:upp_bus_maxV_1})-(\ref{Eq:upp_bus_maxV_2}):
\begin{equation}
    \label{Eq:upp_bus_maxV_1}
    \begin{aligned}
        T_{ni}^{\text{app}} \geq \frac{L_{i}^{\text{app}}}{v^{\text{max}}}, \quad \forall n,i
    \end{aligned}
\end{equation}
\begin{equation}
    \label{Eq:upp_bus_maxV_2}
    \begin{aligned}
        T_{ni}^{\text{dep}} \geq \frac{L_{i}^{\text{dep}}}{v^{\text{max}}}, \quad \forall n,i
    \end{aligned}
\end{equation}

Note that in (\ref{Eq:upp_bus_dep}), intersection delay $d_{ni}$ is tricky to express since the signal cycle index $\hat{k}$ in which bus $n$ would pass intersection $i$ remains unknown. Inspired by~\cite{zengRouteBasedTransitSignal2021}, we introduce an auxiliary binary variable $\theta_{nijk}$, indicating whether bus $n$ would pass intersection $i$ at cycle $k$. Thereafter, $d_{ni}$ can be modeled and converted into a tractable formation with big-M trick, shown as (\ref{Eq:upp_bus_thetasum})-(\ref{Eq:upp_bus_delay_5}).
\[
    \hspace{-6cm} \forall n,i,k, j \in P_{i}^{\text{bus}}:
\]
\begin{equation}
    \label{Eq:upp_bus_thetasum}
    \begin{aligned}
        \sum_{k=1}^{K} \theta_{nijk}=1
    \end{aligned}
\end{equation}
\begin{equation}
    \label{Eq:upp_bus_delay_1}
    \begin{aligned}
        r_{ni} \geq t_{ij,k-1}+g_{ij,k-1}+Y-M(1-\theta_{nijk}),
    \end{aligned}
\end{equation}
\begin{equation}
    \label{Eq:upp_bus_delay_2}
    \begin{aligned}
        r_{ni} \leq t_{ijk}+g_{ijk}+Y+M(1-\theta_{nijk}), 
    \end{aligned}
\end{equation}
\begin{equation}
    \label{Eq:upp_bus_delay_3}
    \begin{aligned}
        \tilde{d}_{ni} \geq t_{ijk}-r_{ni}-M(1-\theta_{nijk})
    \end{aligned}
\end{equation}
\begin{equation}
    \label{Eq:upp_bus_delay_4}
    \begin{aligned}
        \tilde{d}_{ni} \leq t_{ijk}-r_{ni}+M(1-\theta_{nijk})
    \end{aligned}
\end{equation}
\begin{equation}
    \label{Eq:upp_bus_delay_5}
    \begin{aligned}
        d_{ni} \geq \max \{\tilde{d}_{ni}, 0\}
    \end{aligned}
\end{equation}
where $\tilde{d}_{ni}$ is introduced to prevent $d_{ni}$ becoming negative.

In summary, by reformulating the absolute operator in~(\ref{Eq:upp_obj}), we could derive the final problem formulation (\ref{Eq:upp_opt}), which is a mixed integer linear programming (MILP) problem that can be solved efficiently with commercial solvers, such as CPLEX, GUROBI et al.
\begin{equation}
    \label{Eq:upp_opt}
    \begin{aligned}
        &\min \sum_{i=1}^{I} \sum_{j \in P} \sum_{k=1}^{K} w_c(G_{ij}^{\text{opt}}-g_{ijk}) + \sum_{n=1}^{N} \sum_{i=1}^{I} w_b(t_{ni}^{\text{dev}}) \\
        &\text{s.t.} \quad t_{ni}^{\text{dev}} \geq t_{ni}^{\text{arr}}-t_{ni}^{\text{opt}}, \\
        &\quad \quad t_{ni}^{\text{dev}} \geq -(t_{ni}^{\text{arr}}-t_{ni}^{\text{opt}}), \\
        &\quad \quad (\ref{Eq:upp_tls_inK})-(\ref{Eq:upp_bus_maxV_2})
    \end{aligned}
\end{equation}

Notably, although the upper model could output an optimized signal timing plan and bus speed guidance, its performance under stochastic environment is limited due to model mismatch problem. Therefore, only the planned bus arrival time $t^{\text{arr}}_{ni}$ and signal cycle assignment $\theta_{nijk}$ is utilized as a high-level guidance that incorporates multiple signals-buses coordination information into the lower model, where the uncertainties are handled explicitly through stochastic programming.

\subsection{Lower Level: Distributed Stochastic Programming}
\begin{figure}[ht]
    \centering
    \includegraphics[width=0.9\linewidth]{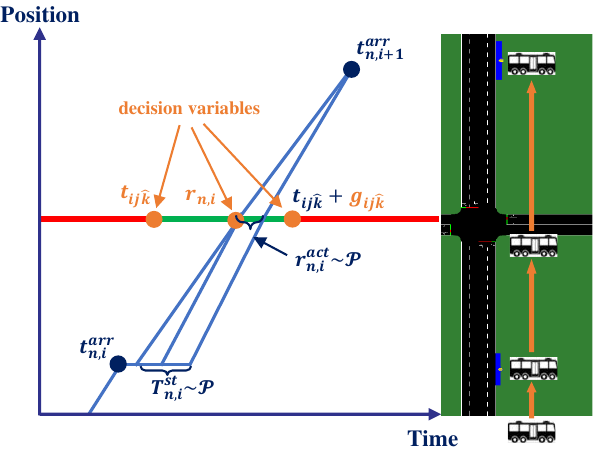}
    \caption{Illustration of stochastic bus operation and decision variables}
    \vspace{-3mm}
    \label{Fig:SP_illu}
    \vspace{-2mm}
\end{figure}

The lower model primarily focuses on addressing the uncertainties of bus movement on each route segment, as illustrated in Fig.~\ref{Fig:SP_illu}. The objective is similar to (\ref{Eq:upp_obj}) but formulated for each intersection individually, with an explicit modeling of stochastic dwell time $T^{\text{st}}$, as shown in (\ref{Eq:low_obj}).
\begin{equation}
    \label{Eq:low_obj}
    \begin{aligned}
        \min J= \sum_{n=1}^{N} w_b\mathbb{E}_{T_{n}^{\text{st}}}[\left| t_{\text{next},n}^{\text{arr}}-t_{\text{next},n}^{\text{upp}} \right|] + \\
        \sum_{j \in P} \sum_{k=1}^{K} w_c(G_{j}^{\text{opt}}-g_{jk})
    \end{aligned}
\end{equation}
where $t_{\text{next},n}^{\text{upp}}$ denotes the planned bus arrival time at the downstream stop of the route segment, which is obtained from the upper model outputs. Other notations follow the same definition as in the upper model.

Similar to the upper model, a mathematical representation of multi-cycle traffic signal timing and bus movement are also required in the lower model. Likewise, the major challenge lies in the modeling of bus movement under stochastic dwell time, since the modeling of multi-cycle signal timing could share the same formulation (\ref{Eq:upp_tls_inK})-(\ref{Eq:upp_tls_yp}) asderived before.

For stochastic bus movement modeling, we first consider the actual arrival time of bus $n$ at the intersection stop line, which is represented by $r_{n}^{\text{act}}$, under stochastic dwell time $T^{\text{st}}$. A piecewise linear function can be derived to formulate $r_{n}^{\text{act}}$, depending on whether bus $n$ could arrive at the intersection as planned arrival time $r_{n}$, without violating the speed limitation constraints~(\ref{Eq:upp_bus_maxV_1})-(\ref{Eq:upp_bus_maxV_2}). The expression of $r_{n}^{\text{act}}$ is shown as:
\begin{equation}
    \label{Eq:low_bus_ract}
    \begin{aligned}
        r_n^{\text{act}} =
        \begin{cases}
            t_n^{\text{arr}} + T_n^{\text{st}} + \frac{l^{\text{app}}}{v^{\max}}, & \text{if } t_n^{\text{arr}} + T_n^{\text{st}} + \frac{L^{\text{app}}}{v^{\max}} > r_n, \\
            r_n, & \text{otherwise}.
        \end{cases}
    \end{aligned}
\end{equation}

Here, $r_{n}$ is taken as a decision variable, representing the impact of bus speed guidance to the probability distribution of $r_{n}^{\text{act}}$, which is a function of random variable $T^{\text{st}}$. With $r_{n}^{\text{act}}$, the actual time that bus $n$ passes the intersection can be further derived, denoted by $\tilde{r}_n^{\text{act}}$, as shown in (\ref{Eq:low_bus_rpass}).
\begin{equation}
    \label{Eq:low_bus_rpass}
    \begin{aligned}
        \tilde{r}_n^{\text{act}} =
        \begin{cases}
            r_n^{\text{act}}, & \text{if } r_n^{\text{act}} \leq t_{j\hat{k}} + g_{j\hat{k}}, \\
            t_{j, \hat{k}+1}, & \text{otherwise}.
        \end{cases}
    \end{aligned}
\end{equation}
where $\hat{k}$ is the assigned signal cycle index for bus $n$ at current intersection, which is determined in $\theta_{nij\hat{k}}$ that output by the upper model. $t_{jk}$ and $g_{jk}$ denote the start time and duration of the bus phase $j$, which are also decision variables of the lower model, along with previous $r_{n}$. This formulation indicates that the bus could either pass the intersection at the middle of green time or have to wait until the red time terminates, depending on the traffic signal state at $r_n^{\text{act}}$.

Subsequently, for the process of bus $n$ approaching next stop $i+1$, we can only consider the late arrival situation, i.e. when bus $n$ cannot arrive at the expected time $t_{\text{next},n}^{\text{upp}}$ scheduled by the upper model because of the speed limit. Otherwise the arrival time deviation can be lowered to zero with the effect of speed adjustment. Thus, the actual bus arrival time at next stop is derived as (\ref{Eq:low_bus_tnext}), along with the reformulation of original objective function (\ref{Eq:low_obj}) into (\ref{Eq:low_obj_new}), where the arrival time deviation is equivalently substituted with arrival lateness.
\begin{equation}
    \label{Eq:low_bus_tnext}
    \begin{aligned}
        t_{\text{next},n}^{\text{arr}}=\tilde{r}_n^{\text{act}}+\frac{L^{\text{dep}}}{v^{\max}}, \quad \forall n
    \end{aligned}
\end{equation}
\begin{equation}
    \label{Eq:low_obj_new}
    \begin{aligned}
        &\min J= \sum_{j \in P} \sum_{k=1}^{K} (G_{j}^{\text{opt}}-g_{jk}) + \sum_{n=1}^{N}  \mathbb{E}_{T_{n}^{\text{st}}}[t_{\text{next},n}^{\text{dev}}]  \\
        &\text{s.t.} \quad t_{\text{next},n}^{\text{dev}} \geq t_{\text{next},n}^{\text{arr}}-t_{\text{next},n}^{\text{upp}}, \\
        &\quad \quad t_{\text{next},n}^{\text{dev}} \geq 0, \quad \forall n
    \end{aligned}
\end{equation}

Thus, the relationship of the objective function (\ref{Eq:low_obj_new}) w.r.t decision variables $r_{n}, t_{j\hat{k}}$ and $g_{j\hat{k}}$ has been constructed by combining (\ref{Eq:low_bus_ract})-(\ref{Eq:low_bus_tnext}), which is reformulated as (\ref{Eq:low_bus_tnext_all}). Here, $e_1$-$e_3$ denotes $t_n^{\text{arr}} + T_n^{\text{st}} + \frac{l^{\text{app}}}{v^{\max}}$, $r_n$ and $t_{j, \hat{k}} + g_{j, \hat{k}}$ respectively.
\begin{equation}
    \label{Eq:low_bus_tnext_all}
    \begin{aligned}
        t_{n, \text{next}}^{\text{arr}} =
        \begin{cases}
            t_n^{\text{arr}} + T_n^{\text{st}} + \frac{L^{\text{app}} + L^{\text{dep}}}{v^{\max}}, 
            & \text{if } e_1 > e_2 \text{ and } e_1 \leq e_3, \\

            r_n + \frac{L^{\text{dep}}}{v^{\max}}, 
            & \text{if } e_1 < e_2 \text{ and } e_2 \leq e_3, \\

            t_{j, \hat{k}+1} + \frac{L^{\text{dep}}}{v^{\max}}, 
            & \text{otherwise}.
        \end{cases}
    \end{aligned}
\end{equation}

To convert~(\ref{Eq:low_bus_tnext_all}) into a solver-tractable formation, binary auxiliary variables $\beta_{n}^{(1)}$-$\beta_{n}^{(3)}$ are introduced to construct a set of switching constraints, as shown in (\ref{Eq:low_bus_tnext_all_})-(\ref{Eq:low_bus_tnext_all_9}).
\begin{equation}
    \label{Eq:low_bus_tnext_all_}
    \begin{aligned}
        t_{n,\text{next}}^{\text{arr}} = \beta_n^{(1)}(t_n^{\text{arr}} + T_n^{\text{st}} + \frac{L^{\text{app}}}{v^{\max}}) + \beta_n^{(2)}(r_n) + \\
        \beta_n^{(3)}(t_{j,\hat{k}+1}) + \frac{L^{\text{dep}}}{v^{\max}}
    \end{aligned}
\end{equation}
\begin{equation}
    \label{Eq:low_bus_tnext_all_0}
    \begin{aligned}
        \beta_n^{(1)} + \beta_n^{(2)} + \beta_n^{(3)} = 1
    \end{aligned}
\end{equation}
\begin{equation}
    \label{Eq:low_bus_tnext_all_1}
    \begin{aligned}
        t_n^{\text{arr}} + T_n^{\text{st}} + \frac{L^{\text{app}}}{v^{\max}} > r_n - M (1 - \beta_n^{(1)}), \quad \forall n
    \end{aligned}
\end{equation}
\begin{equation}
    \label{Eq:low_bus_tnext_all_2}
    \begin{aligned}
        t_n^{\text{arr}} + T_n^{\text{st}} + \frac{L^{\text{app}}}{v^{\max}} \leq t_{j\hat{k}} + g_{j\hat{k}} + M (1 - \beta_n^{(1)})
    \end{aligned}
\end{equation}
\begin{equation}
    \label{Eq:low_bus_tnext_all_3}
    \begin{aligned}
        t_n^{\text{arr}} + T_n^{\text{st}} + \frac{L^{\text{app}}}{v^{\max}} \leq r_n + M (1 - \beta_n^{(2)})
    \end{aligned}
\end{equation}
\begin{equation}
    \label{Eq:low_bus_tnext_all_4}
    \begin{aligned}
        r_n \leq t_{j\hat{k}} + g_{j\hat{k}} + M (1 - \beta_n^{(2)})
    \end{aligned}
\end{equation}
\begin{equation}
    \label{Eq:low_bus_tnext_all_5}
    \begin{aligned}
        t_n^{\text{arr}} + T_n^{\text{st}} + \frac{L^{\text{app}}}{v^{\max}} > t_{j\hat{k}} + g_{j\hat{k}} - M (1 - \beta_n^{(3a)})
    \end{aligned}
\end{equation}
\begin{equation}
    \label{Eq:low_bus_tnext_all_6}
    \begin{aligned}
        r_n > t_{j\hat{k}} + g_{j\hat{k}} - M (1 - \beta_n^{(3b)})
    \end{aligned}
\end{equation}
\begin{equation}
    \label{Eq:low_bus_tnext_all_7}
    \begin{aligned}
        \beta_n^{(3)} \geq \beta_n^{(3a)}
    \end{aligned}
\end{equation}
\begin{equation}
    \label{Eq:low_bus_tnext_all_8}
    \begin{aligned}
        \beta_n^{(3)} \geq \beta_n^{(3b)}
    \end{aligned}
\end{equation}
\begin{equation}
    \label{Eq:low_bus_tnext_all_9}
    \begin{aligned}
        \beta_n^{(3)} \leq \beta_n^{(3a)} + \beta_n^{(3b)}
    \end{aligned}
\end{equation}
where special tricks are applied in (\ref{Eq:low_bus_tnext_all_5})-(\ref{Eq:low_bus_tnext_all_9}) to equivalently express the 'or' operator in the last branch of piecewise function~(\ref{Eq:low_bus_tnext_all}).

In summary, the whole formulation of the lower model is written as~(\ref{Eq:low_opt}).
\begin{equation}
    \label{Eq:low_opt}
    \begin{aligned}
        &\min \sum_{j \in P} \sum_{k=1}^{K} (G_{j}^{\text{opt}}-g_{jk}) + \sum_{n=1}^{N}  \mathbb{E}_{T_{n}^{\text{st}}}[t_{\text{next},n}^{\text{dev}}]  \\
        &\text{s.t.} \quad (\ref{Eq:low_obj_new}),(\ref{Eq:low_bus_tnext_all_0})-(\ref{Eq:low_bus_tnext_all_9})
    \end{aligned}
\end{equation}

Adopting the Sample Average Approximation (SAA) method, the expectation term in (\ref{Eq:low_opt}) is estimated by the average of multiple $t_{\text{next},n}^{\text{dev}}$, each calculated by one $T_{n}^{\text{st}}$ sample. Hence, (\ref{Eq:low_opt}) can be equivalently transformed into an MILP problem and solved efficiently by commercial solvers. Notably, since each lower model could run independently with the guidance of upper model outputs, those lower models can be solved in parallel, thus significantly improves computational efficiency and scalability of the algorithm.

\section{Simulation and Results}

\begin{figure}[ht]
    \vspace{-2mm}
    \centering
    \includegraphics[width=\linewidth]{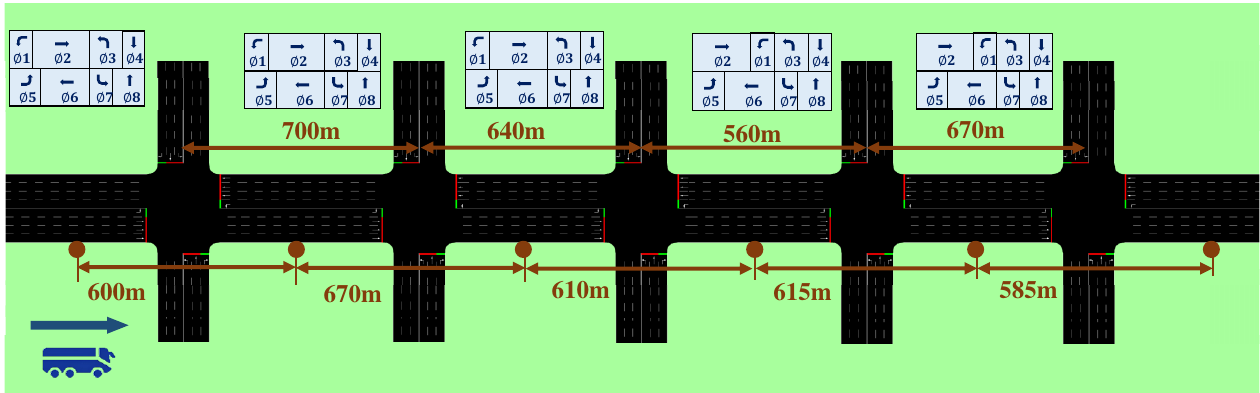}
    \caption{Test corridor with five coordinated intersections.}
    \vspace{-3mm}
    \label{Fig:simu_setup}
\end{figure}

A test corridor with five consecutive signalized intersections is constructed in SUMO to validate the proposed strategy, as illustrated in Fig.~\ref{Fig:simu_setup}. Both main-street directions, i.e. westbound and eastbound direction, consist of one exclusive left-turn lane, two through lanes and one dedicated bus lane, while each cross-street direction has one left-turn lane, one through lane and one right-turn lane. The saturation flow rates is set aligned with~\cite{maMultistageStochasticProgram2016a}. The signal phases are indexed following the National Electrical Manufacturers Association (NEMA) standard. The background signal timing plan is optimized offline using MAXBAND algorithm~\cite{littleVersatileProgramSetting}.

\begin{table}[h]
    \centering
    \small
    \caption{Traffic volume and signal timing setup}
    \vspace{-2mm}
    \resizebox{\linewidth}{!}{ 
    \begin{tabular}{ccccccccc}
        \toprule
        \multicolumn{9}{c}{\textbf{Intersection 1 (offset = 0)}} \\
        \midrule
        Phase & $\phi_1$ & $\phi_2$ & $\phi_3$ & $\phi_4$ & $\phi_5$ & $\phi_6$ & $\phi_7$ & $\phi_8$ \\
        \midrule
        Vol. (vph) & 157 & 1118 & 248 & 169 & 244 & 720 & 34 & 135 \\
        Splits (s) & 14 & 48 & 23 & 15 & 25 & 37 & 12 & 26 \\
        \midrule
        \multicolumn{9}{c}{\textbf{Intersection 2 (offset = 44)}} \\
        \midrule
        Phase & $\phi_1$ & $\phi_2$ & $\phi_3$ & $\phi_4$ & $\phi_5$ & $\phi_6$ & $\phi_7$ & $\phi_8$ \\
        \midrule
        Vol. (vph) & 172 & 951 & 90 & 311 & 207 & 789 & 89 & 314 \\
        Splits (s) & 17 & 45 & 12 & 26 & 22 & 40 & 12 & 26 \\
        \midrule
        \multicolumn{9}{c}{\textbf{Intersection 3 (offset = 54)}} \\
        \midrule
        Phase & $\phi_1$ & $\phi_2$ & $\phi_3$ & $\phi_4$ & $\phi_5$ & $\phi_6$ & $\phi_7$ & $\phi_8$ \\
        \midrule
        Vol. (vph) & 186 & 846 & 97 & 435 & 184 & 853 & 124 & 339 \\
        Splits (s) & 17 & 37 & 12 & 34 & 17 & 37 & 13 & 33 \\
        \midrule
        \multicolumn{9}{c}{\textbf{Intersection 4 (offset = 30)}} \\
        \midrule
        Phase & $\phi_2$ & $\phi_1$ & $\phi_3$ & $\phi_4$ & $\phi_5$ & $\phi_6$ & $\phi_7$ & $\phi_8$ \\
        \midrule
        Vol. (vph) & 795 & 216 & 65 & 291 & 173 & 989 & 58 & 327 \\
        Splits (s) & 40 & 23 & 12 & 25 & 17 & 46 & 12 & 25 \\
        \midrule
        \multicolumn{9}{c}{\textbf{Intersection 5 (offset = 43)}} \\
        \midrule
        Phase & $\phi_2$ & $\phi_1$ & $\phi_3$ & $\phi_4$ & $\phi_5$ & $\phi_6$ & $\phi_7$ & $\phi_8$ \\
        \midrule
        Vol. (vph) & 699 & 244 & 104 & 301 & 152 & 1118 & 116 & 270 \\
        Splits (s) & 36 & 27 & 12 & 25 & 14 & 49 & 12 & 25 \\
        \bottomrule
    \end{tabular}
    }
    \vspace{-3mm}
    \label{Tab:sim_setup}
\end{table}

Three test scenarios are designed to present comprehensive evaluation of proposed strategy, corresponding to different traffic demand. Here, we adopt real-world peak-hour traffic demand data collected in~\cite{ZGGL202310020} multiplied by a scaling coefficient 0.9/0.8/0.7 to generate the three test scenarios, i.e. high/medium/low-volume scenario. The corresponding max v/c ratios are 0.85/0.77/0.67, representing near-saturation/medium congestion/low congestion traffic condition, respectively. The detailed settings of high-volume scenario is presented in Table~\ref{Tab:sim_setup}.

A bus route segment consisting of 6 stops is set up on the test corridor, as illustrated in Fig.~\ref{Fig:simu_setup}. Buses are running westbound with a maximum speed of 12 m/s (43 kph), about 10 kph lower than passenger cars free-flow speed, which is set as 15 m/s (54 kph). A pre-defined bus timetable is designed to determine scheduled arrival times for each bus at each stop. The bus departure frequency is set as 2 minutes to simulate the arrival of a busy bus route or several overlapping common bus routes. The expected average speed for each bus is set as 10 m/s (36 kph), which is a relatively high value to highlight the effect of TSP. Under the configuration above, 100\% of buses are late at the last stop under the effect of intersection delay. The stochastic bus dwell time at each stop follows a uniform distribution $\mathbb{U}[15, 35]$s, which is consistent with the reference values in~\cite{daganzoHeadwaybasedApproachEliminate2009}. The whole simulation ends at 3600s, including the trajectory of about 30 buses.

An enhanced version of R-TSP algorithm proposed in~\cite{zengRouteBasedTransitSignal2021}, named RTSP-SA, is selected as the comparative method, which is a route-level integrated signal priority and speed control strategy based on deterministic optimization. Considering the stochastic nature of bus dwell time, we conduct 30 rounds of simulation repeatedly for each scenario and select average performance to diminish the effect of occasional abnormally bad or good results. The sample size of SAA algorithm is selected as 50, under which a satisfactory result could be achieved with an average computational time less than 2s, which is acceptable for real-time applications.

\begin{table}[h]
    \caption{Simulation results of schedule adherence}
    \centering
    \resizebox{\linewidth}{!}{ 
    \begin{tabular}{llcccccc}
        \toprule
        & \multirow{2}{*}{Vol.} & \multicolumn{2}{c}{Schedule Dev.(s)} & \multicolumn{2}{c}{Headway SD.(s)} & \multicolumn{2}{c}{Punctual Rate} \\
        \cmidrule(lr){3-4} \cmidrule(lr){5-6} \cmidrule(lr){7-8}
        & & Avg. & \% & Avg. & \% & Avg. & \% \\
        \midrule
        \multirow{3}{*}{Blank} 
        & low & 47.1 & - & 51.2 & - & 47.9\% & - \\
        & med & 46.5 & - & 52.4 & - & 47.9\% & - \\
        & high & 47.9 & - & 52.0 & - & 48.0\% & - \\
        \midrule
        \multirow{3}{*}{RTSP-SA} 
        & low & 20.7 & -56.1\% & 35.2 & -31.3\% & 74.6\% & +55.7\% \\
        & med & 25.0 & -46.2\% & 38.1 & -27.3\% & 68.7\% & +43.4\% \\
        & high & 27.9 & -41.8\% & 40.0 & -23.1\% & 67.4\% & +40.4\% \\
        \midrule
        \multirow{3}{*}{HierTSP-SA} 
        & low & \textbf{4.9} & \textbf{-89.6\%} & \textbf{9.5} & \textbf{-81.5\%} & \textbf{99.6\%} & \textbf{+107.9\%} \\
        & med & \textbf{6.9} & \textbf{-85.2\%} & \textbf{13.3} & \textbf{-74.6\%} & \textbf{99.2\%} & \textbf{+107.1\%} \\
        & high & \textbf{6.9} & \textbf{-85.6\%} & \textbf{13.9} & \textbf{-73.3\%} & \textbf{99.2\%} & \textbf{+106.7\%} \\
        \bottomrule
    \end{tabular}
    }
    \label{Tab:busPI}
\end{table}

Table~\ref{Tab:busPI} displays the bus schedule adherence performance of three methods, i.e. no control (Blank), baseline method (RTSP-SA) and our proposed method (HierTSP-SA), under different traffic demand levels. The selected performance indices include bus schedule deviation, bus time headway standard deviation and bus punctual rate. The punctual rate here is defined as the percentage of bus arrivals that deviates less than 30 seconds from timetable. All these indices are averaged across all bus arrivals at 6 stops. As shown in Table~\ref{Tab:busPI}, HierTSP-SA could effectively mitigate schedule deviations by 85.2\%-89.6\% under three scenarios, improving the punctual rate from 47.9\% to 99.2\%-99.6\%, which significantly outperforms RTSP-SA. Furthermore, RTSP-SA only achieves a minor improvement of 23.1\%-31.3\% on bus time headway equalization, which means the time interval of two bus arrivals fluctuates severely under stochastic disturbances. In contrast, HierTSP-SA could maintain a stable time headway that fluctuates less than 13.9s in average, achieving a promotion of 73.3\%-81.5\%. In addition, as the traffic demand increases, the performance of HierTSP-SA shows less variation compared with RTSP-SA, indicating a reliable schedule adherence promotion under various traffic conditions.

\begin{figure}[hb]
    \vspace{-2mm}
    \centering
    \includegraphics[width=0.9\linewidth]{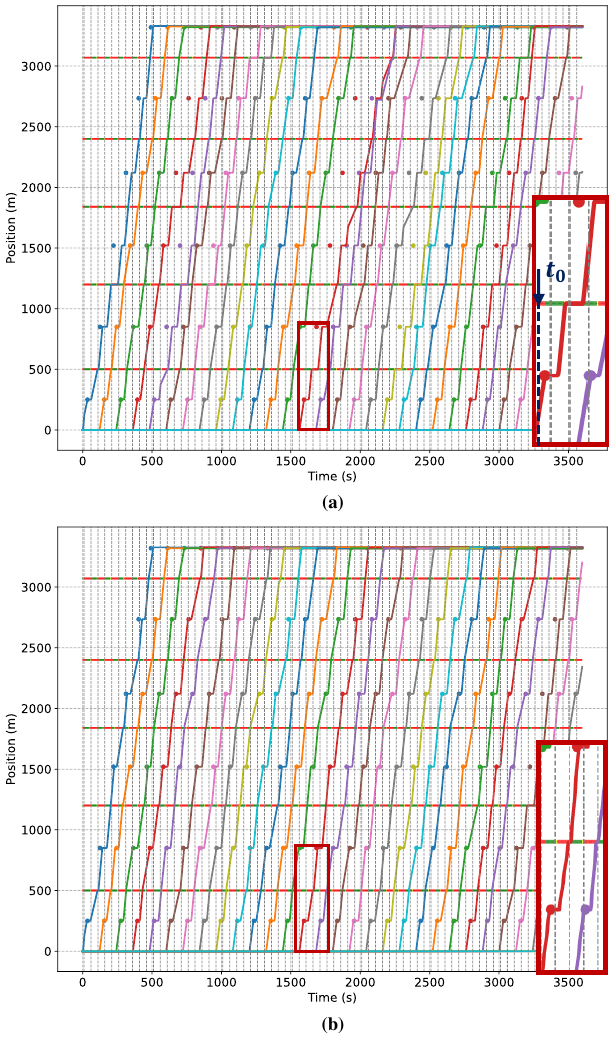}
    \vspace{-3mm}
    \caption{Trajectories of buses in a single medium-volume test scenario}
    \label{Fig:traj}
    \vspace{-3mm}
\end{figure}

To further understand how HierTSP-SA could maintain robust performance under stochastic disturbances, the bus trajectories of two control strategies are compared and analyzed, as shown in Fig.~\ref{Fig:traj}. The color of horizontal lines represents corresponding traffic signal states, i.e. green, yellow or red. As demonstrated in Fig.~\ref{Fig:traj}(a), the irregular bus arrival of RTSP-SA mainly attribute to a typical failure situation shown in the red box. When the signal timing is optimized in $t_0$, the stochastic bus dwell time is still unrevealed, thus is set as a point estimation (i.e. mean value) in RTSP-SA due to its deterministic nature. However, if the actual dwell time is highly deviated from the point estimation, the bus can not catch up with the initial speed guidance and pass the intersection at the allocated signal cycle, thus resulting in an infeasible plan. In comparison, HierTSP-SA estimate the dwell time with probability distribution to generate robust signal timing plan and speed guidance in advance, thereby eliminates the model mismatch problem and effectively lower the risk of the failure situation.


\begin{figure}[hb]
    \vspace{-2mm}
    \centering
    \includegraphics[width=\linewidth]{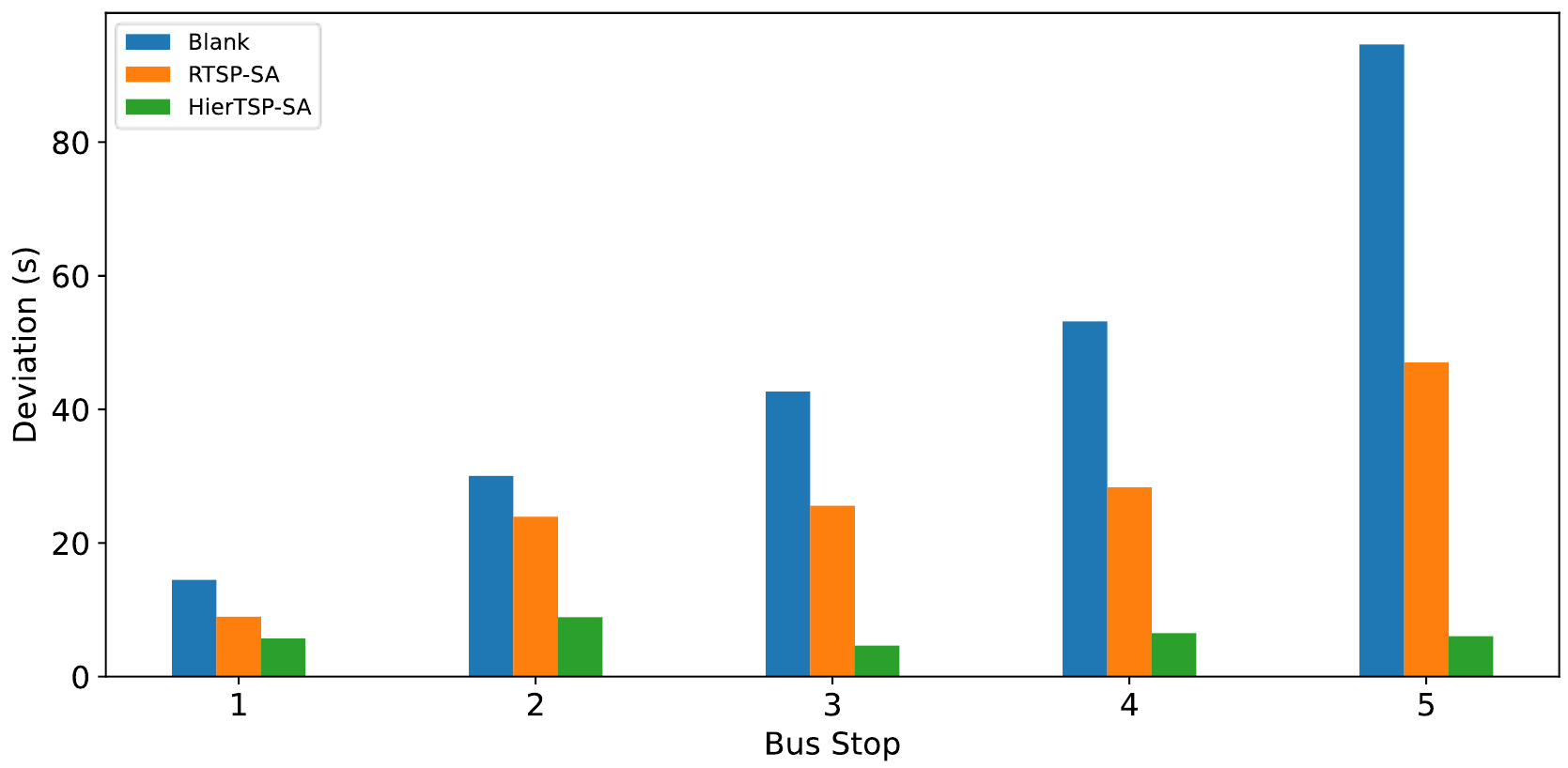}
    \vspace{-3mm}
    \caption{Variation of bus arrival time deviation towards downstream intersections}
    \label{Fig:busArrTimeDev}
    \vspace{-3mm}
\end{figure}

Fig.~\ref{Fig:busArrTimeDev} further validates the bus schedule adherence performance across consecutive bus stops. The horizontal axis represents the bus stop number, which is indexed from upstream to downstream. The result indicates that when no control or RTSP-SA is applied in stochastic environments, the schedule deviation tends to diverge as the bus moving downstream, which could easily resulting in bus bunching as the bus route extends. In contrast, the performance of HierTSP-SA maintains nearly invariant along the whole route, implying stability and high robustness to randomness.

\begin{table}[hb]
    \caption{Simulation results of traffic impact}
    \centering
    \resizebox{\linewidth}{!}{ 
    \begin{tabular}{llcccccc}
        \toprule
        & \multirow{2}{*}{Vol.} & \multicolumn{2}{c}{Delay (s)} & \multicolumn{2}{c}{Max Queue (veh)} & \multicolumn{2}{c}{Stop} \\
        \cmidrule(lr){3-4} \cmidrule(lr){5-6} \cmidrule(lr){7-8}
        & & Avg. & \% & Avg. & \% & Avg. & \% \\
        \midrule
        \multirow{3}{*}{Blank} 
        & low & 24.5 & - & 4.7 & - & 0.606 & - \\
        & med & 26.2 & - & 5.6 & - & 0.641 & - \\
        & high & 28.8 & - & 6.1 & - & 0.675 & - \\
        \midrule
        \multirow{3}{*}{RTSP-SA} 
        & low & 24.6 & +0.4\% & 4.6 & -2.1\% & 0.603 & -0.5\% \\
        & med & 26.2 & $\pm$ 0\% & 5.5 & -1.8\% & 0.636 & -0.8\% \\
        & high & 29.3 & +1.7\% & 6.1 & $\pm$ 0\% & 0.676 & +0.2\% \\
        \midrule
        \multirow{3}{*}{HierTSP-SA} 
        & low & \textbf{24.9} & \textbf{+1.6\%} & \textbf{4.8} & \textbf{+2.1\%} & \textbf{0.609} & \textbf{+0.5\%} \\
        & med & \textbf{26.4} & \textbf{+0.8\%} & \textbf{5.8} & \textbf{+3.6\%} & \textbf{0.641} & \textbf{$\pm$ 0\%} \\
        & high & \textbf{30.3} & \textbf{+5.2\%} & \textbf{6.5} & \textbf{+6.6\%} & \textbf{0.695} & \textbf{+3.0\%} \\
        \bottomrule
    \end{tabular}
    }
    \label{Tab:traff_PI}
\end{table}

The traffic impacts of proposed method under various traffic demands are presented in Table~\ref{Tab:traff_PI}, which are evaluated by average passenger car delay, average maximum queue length and average passenger car stops. Note that under low and medium volume scenarios, our approach only introduces an extra delay of 0.8\%-1.6\%, which is almost negligible to the overall traffic efficiency. While in high volume condition, a higher 5.2\% delay is introduced since the priority resources are extremely limited for reliable bus movement, given the near-saturated traffic condition. However, an increase of 5\% is still acceptable and aligns with the performance of current algorithms such as R-TSP, as given in~\cite{zengRouteBasedTransitSignal2021}. In summary, under various traffic conditions, our proposed method could effectively promote bus schedule adherence with a limited increase of car delay, meanwhile maintaining high robustness against stochastic disturbances.


\section{Conclusion}
  In this study, a robust integrated signal priority and bus speed control strategy based on hierarchical stochastic optimization is proposed, to promote bus schedule adherence in signalized arterial. The primary conclusions are summarized as follows.

  \begin{itemize}
    \item [1)] In the arterial level, the integrated optimization process is formulated as a deterministic MILP problem which schedules multi-cycle signal plan and multi-bus speed in a global perspective. As revealed by simulation results, route-level bus schedule adherence is achieved with a limited car delay increase of 0.8\%-5.2\% on various traffic condition including near-saturated scenario, which demonstrates the effective coordination across intersections and buses.
    \item [2)] In the intersection level, stochastic programming is adopted to explicit handles stochastic disturbances without reliance on abrupt signal timing variation. Simulation results demonstrate that our method significantly outperforms deterministic optimization based methods under stochastic environments, with an improvement ranging from 85.2\% to 89.6\% on bus schedule adherence, and the deviation maintains converged as buses moving downstream the route, validating the effectiveness and robustness of the proposed approach.
  \end{itemize}

  For future research, we intend to extend the proposed framework into multi-route scenarios. Furthermore, a field experiment will be conducted to further validate the effectiveness of proposed strategy in real-world.

\bibliographystyle{IEEEtran}
\bibliography{./bibfile/bibConf.bib}

\end{document}